\documentclass[trackchanges, twocolumn]{aastex7}
\usepackage{amsmath}

\begin{document}

\title{Comparison of Different Methods to Determine the Distance to LMC}

\author[orcid=0009-0005-4685-695X,sname='Sánchez Medina', gname='Juan José']{Juan José Sánchez Medina}
\affiliation{Colegio Andino de Tunja}
\email[show]{jjsanchezmedina@gmail.com}  








\begin{abstract}

This research paper aims to compare different methods for calculating the distance to the Large Magellanic Cloud (\textit{LMC}). 
The distance, $d_{LMC}$, is determined using stellar parallax, variable stars (RR Lyrae and Classical Cepheids), redshift, and celestial mechanics, from which the systematic and standard errors are calculated. 
After analyzing each method, the final distance is obtained as $d_{LMC} = 50.4802 \pm 0.0638_{\text{std}}$ Kpc, differing by $+0.5102$ Kpc from the currently most accepted value of $d_{LMC} = 49.97$ Kpc \citep{26_Pietrzyński_2014}. 
In this paper, the value of $d_{LMC}$ was derived by combining the distances determined from RR Lyrae and Classical Cepheid variable stars, celestial mechanics and parallax.

\end{abstract}



\section{Introduction} \label{sec:introduction}

The development of the notion of our location in the universe and the mapping of the universe from the local system to our location in groups of galaxies has been a recurring topic of research in recent decades in astrophysics and cosmology. This field of research has allowed us to establish relationships and patterns in how our universe is organized, calibrate stellar measurements and parameters, recognize the distribution of matter in it; as well as, create new models that allow supporting precision in cosmological calculations such as the Hubble-Lemaitre model, thanks to the fact that we can observe patterns in the dilation and expansion of the universe; as well as trace how the universe looked at its beginning by comparing different measurements and identifying trends \citep{31_Freedman_2019, 22_Brescher_2023}.

Measuring distances in space has been a topic of great interest since the beginning of astronomy, and geometric, optical, physical and cosmological methods have been developed over more than 2000 years; methods that differ in their levels of precision and error depending on the distance and scale in question \citep{23_Manahel_2013}.

When we talk about the local group, we find that the Milky Way is accompanied by more than 60 satellite galaxies \citep{40_Bovill_2011}, where the Small Magellanic Cloud (SMC) and the Large Magellanic Cloud (LMC) stand out. The latter is a dwarf galaxy of type SBm\footnote{Barred spiral galaxy.}, located in the constellation of Dorado (Dor) and Mensa (Men), with an angular size of approximately $322.8' \times 274.7'$ and an apparent visual magnitude $m_{v} = 0.13 $ \citep{SIMBAD,NASA/IPAC}.

On the other hand, in the distant future \textit{LMC} will join the Milky Way \citep{30_Cautun_2018}; this future event gains importance since it will allow us to understand the interactions between galaxies that occur in the universe, as well as the possible victims and the outcome of this collision.

The Magellanic Clouds (MC) have been of great help as a research area due to their relative proximity and their characteristics, which has allowed the development of multiple investigations. In the case of the Large Magellanic Cloud, we find highly active areas in star formation, with a high number of young, massive and hot stars; the Tarantula Nebula (NGC\footnote{From the English, New General Catalogue.}2070), the Bean Nebula (NGC1763), NGC1760, NGC1773 or NGC1769; stand out for being the most visible emission nebulae in LMC \citep{Stellarium_2021}. Among other objects with high scientific value in the LMC we have the presence of more than 3050 star clusters, more than 24,000 variable stars of the RR Lyrae type, more than 3300 classical Cepheids and 52 supernova remnants (SNR\footnote{From English, Super Nova Remnant.}) \citep{20Meschin_2011}.

The Large Magellanic Cloud \textit{LMC} is a region of high interest by the scientific community, which is used as a calibration of multiple procedures and calculations, which directly depend on the distance to this \citep{31_Freedman_2019}. For this reason, for decades new methods have been developed and perfected that allow measuring the distance to this satellite galaxy; Each time, new technologies are being used as they become available, such as the Hubble Space Telescope (HST), the James Webb Space Telescope (JWST), the space probe (GAIA) and ground-based observatories such as the European Southern Observatory (ESO).

In this research, the comparison of six methods for measuring distances is used in order to find a value with greater precision. The methods used to determine the distance are: stellar parallax; variable stars, RR Lyrae and Cepheids; eclipsing binaries; celestial mechanics; redshift and the use of the supernova SN1987a.

\section{Background} \label{sec:background}
Centuries ago, philosophers, mathematicians and scientists such as Johannes Kepler and Isaac Newton established the foundations of celestial mechanics by introducing the concept of forces at a distance and thus explaining the movement of the planets and the forces involved in the process; which depend directly on the distance \citep{Newton_1687}. This distance allows us to convert the observable properties of planets, stars, galaxies, black holes, novae or supernovae into physical quantities \citep{36_Blakeslee_2021}.

\begin{figure}[h!]
\centering
\caption{Graphic representation of the Local Group it represents the galaxies of the Local Group (LG) highlighting the Large and Small Magellanic Clouds in red and the galactic center in yellow. Considering the scales of the figure, it can be approximated that the Earth is also located in the yellow dot. A Python script was used to create this figure\footnote{See section \ref{sec:appendix}.}}
\includegraphics[width=8cm]{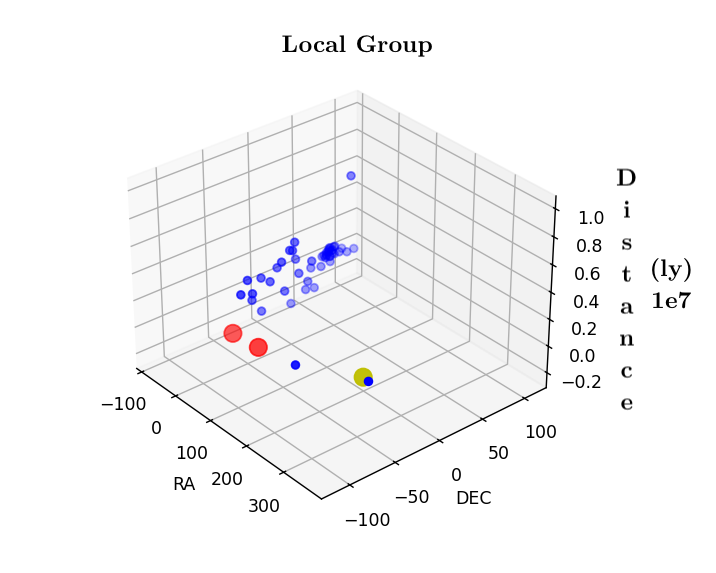}\\
\label{fig:LocalGroup}
\end{figure}

The solar system, located in the Orion arm \citep{8_Croswell_2021}., is located approximately $8.1$ Kpc\footnote{Kiloparsec, $1 \times 10^{3}$ pc.} away from the galactic center \citep{3_Abuter_2019}. Likewise, the Milky Way is located in the so-called \textit{Local Group} sector, determined by a spherical region of approximately 3 Mpc\footnote{Megaparsec, $1 \times 10^{6}$ pc.} in diameter where more than 35 galaxies are found; this relative proximity to the galaxies belonging to the Local Group allows us to resolve individual stars to later perform photometric studies which will be relevant in the present research \citep{37_Bergh_1999, 20Meschin_2011}. As can be seen in figure \ref{fig:LocalGroup}, the proximity to the Magellanic Clouds is significant and provides a first glimpse of the local group in which the Milky Way is located, and therefore, the Solar System.

Additionally, the Large Magellanic Cloud and the Small Magellanic Cloud are the fourth and sixth largest galaxies in the local group, respectively; taking into account their actual diameter\footnote{Diameter referring to the length dimension.} \citep{SIMBAD}.

The Large Magellanic Cloud is the largest satellite galaxy of the Milky Way, estimated to have a mass of approximately $1.38 \times 10^{11} M_{\odot}$ \citep{46_Erkal_2019}., approximately $\frac{1}{9}$ the mass of the Milky Way. Furthermore, as mentioned above, thanks to the research of \citet{30_Cautun_2018} it is estimated that in approximately $2.4 ^{+1.2}_{-0.8}$ Gyrs
the LMC will collide with the Milky Way, thus producing gravitational nudges to star systems along with increased activity in the galactic nucleus (AGN\footnote{From the English, Active Galactic Nucleus}). The significance of this collision is that it will take place long before the collision of the Andromeda galaxy (M\footnote{Refers to the identifier given to the Messier catalogue: M-NN}31) and its satellite galaxies (M32 and M110).

There are currently multiple methods for determining distances in space; each is used according to the scale of the investigation, in the sense of the distance and object being studied; this is called the cosmic distance ladder, according to \citet{47_Reiche_2022} this can be represented according to table \ref{tab:distanceladder}.  

However, these methods can be combined to compare the errors of each one or even to obtain an even more precise measurement, as demonstrated by \citet{14_Clementini_2003} when he compared the errors of 25 different distance determination methods.

\begin{deluxetable*}{lll}
\tablewidth{0pt}
\tablecaption{Cosmic stellar staircase \label{tab:distanceladder}}
\tablehead{
\colhead{Method} & \colhead{Effective Distance} & \colhead{Objects}
}
\startdata
Radar & $10^{-4}$ ly & Solar System \\
Parallax & $10^{2}$ ly & Nearby Stars \\
MS Adjustment & $10^{5}$ ly & Milky Way \\
Cepheids & $10^{7}$ ly & Nearby Galaxies \\
Tully-Fisher & $10^{10}$ ly & Galaxy Clusters \\
Hubble-Lemaitre Law & $>$ $10^{10}$ ly & Redshift $z \geq 10$
\enddata
\tablecomments{This table represents the cosmic distance ladder, where different methods are used to measure distances to astronomical objects at varying scales.}
\end{deluxetable*}

In parallel, with the development of new technologies and better tools, measurements like this one are becoming increasingly popular. Space telescopes and observatories such as JWST, focused on the infrared spectrum; GAIA\footnote{Space Survey led by ESA.}, an ESA observatory focused on astrometry, spectrometry and photometry of the celestial vault; Euclid, which will mainly be responsible for observing the visible spectrum, near-infrared spectrometry and photometry; Chandra, NASA's\footnote{National Aeronautics and Space Administration} X-ray observatory; together with large ground-based observatories such as the VLT\footnote{Very Large Telescope.}, the Gran Telescopio Canarias or future projects such as the ELT\footnote{Extremely Large Telescope}; allow us to resolve objects at an unprecedented resolution thanks in part to new technologies such as adaptive optics; In this way, we can see what the future panorama of astronomy will be, where current processes, research and projects evolve with it and provide us with new tools to continue investigating \citep{48_Gardner_2006, 49_Vernin_2011, 50_Racca_2016, 51_Weisskopf_2002, 52_Scholler_2007}.

Due to the importance of the value of the distance to the Large Magellanic Cloud for astronomy and cosmology, this research seeks to integrate multiple methods for determining distances in space to obtain an accurate value for this satellite galaxy.

\section{Justification} \label{sec:justification}

The universe is a laboratory for astronomers, which has always amazed humanity, and is in constant change and evolution. Astronomy was born with the Egyptian and Mesopotamian civilizations with the need to understand and predict certain phenomena that take place on Earth; from which, later, calendars would emerge that would be fundamental in the evolution of humanity itself as a civilization due to the development of agriculture; in addition, astronomical phenomena such as eclipses, seasons, lunar phases or the construction of sundials would soon be analyzed \citep{54_Valenzuela_2010}. These events gave rise to astronomy as a science that would soon only evolve.

The Milky Way Galaxy, home to the solar system, is a spiral galaxy located approximately $8.1$ Kpc from the galactic center \citep{55_Xu_2023, 3_Abuter_2019}. It has 11 classical satellite galaxies \citep{56_Sawala_2023}, of which we will focus on the Large Magellanic Cloud (LMC). The LMC has records dating back more than 1000 years, which gets its name from the circumnavigation of the Earth in 1519-1522 by the expedition led by Ferdinand Magellan; this is a spiral satellite galaxy of the Milky Way that has been studied since then \citep{20Meschin_2011}. The LMC is a region of study that, together with the SMC, has allowed discoveries to be made regarding star formation, colour index relationships, calibration of the Hubble constant or the famous period-luminosity (PL) relationship \citep{57_ESO_2019, Leavitt_1912, 58_Freedman_2012}.

Measuring distances in space became relevant from the first models of the solar system that were proposed. Later, physicists such as Isaac Newton and astronomers such as Johannes Kepler related the distance to their theories and laws; in the case of Kepler, he developed all his work using the contributions that Nicolaus Copernicus previously made about the distance to the planets using geometric methods. \citep{Kepler_1964,Copernico_1965}.

In the case of the Large Magellanic Cloud, interest in determining its distance has been present since its discovery and has materialized since the last century. Figure \ref{fig:distanciascrono} shows the distance determined by the investigations of \citet{4_Alves_2004, 13_Panagia_1999, 32_McCall_1993, LMCCrono_1, LMCCrono_2, LMCCrono_3, LMCCrono_4, LMCCrono_5, LMCCrono_6, LMCCrono_7, LMCCrono_8, LMCCrono_9, LMCCrono_10, LMCCrono_11, LMCCrono_12} at different times.

\begin{figure}[h!]
\caption{Distances to the LMC over time. This figure represents the different distances to the LMC according to the present research compilation. The first investigations did not have the error estimate.}
\includegraphics[width=8.5cm]{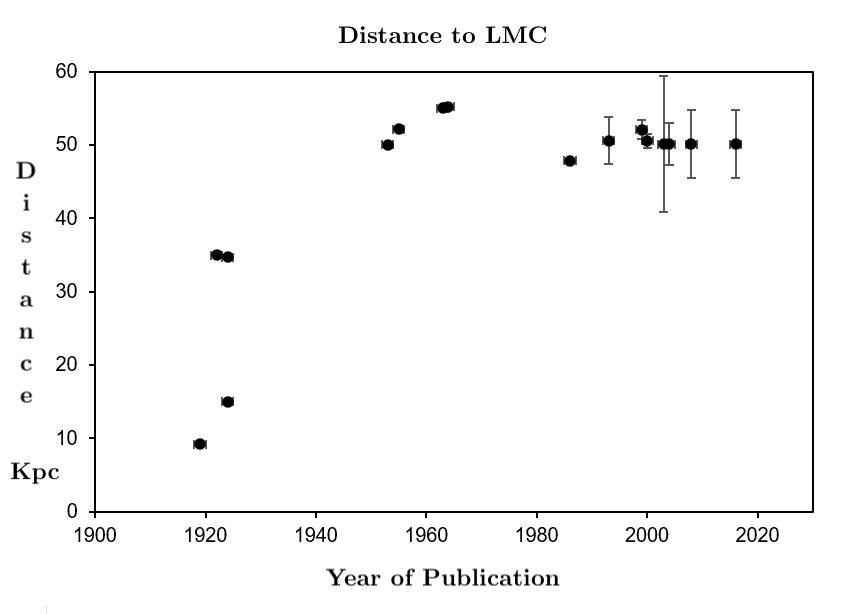}\\
\label{fig:distanceschrono}
\end{figure}

As shown in figure \ref{fig:distanceschrono}, work has been done over the past 100 years to determine this distance. It is also evident that recently, the deviations in the measurement of the distance to the LMC have been considerably reduced. For the reasons stated above, this research will use seven different methods in order to obtain an average of these and arrive at a result with a remarkable degree of precision to the LMC.

\section{Scope} \label{sec:scope}

The main scope of this paper is to compare different methods for determining distances to the Large Magellanic Cloud.

\subsection{Specific Objectives}
\begin{itemize}
    \item Set the different errors for each of the methods used.
    \item Determine the distance to LMC using each method.
    \item Calculate the distance to LMC using the most reliable methods.
\end{itemize}

\section{Theoretical Framework} \label{sec:framework}
\subsection{Conceptual Reference}
\subsubsection{Local Group}
The Local Group is a region with imaginary boundaries around the Milky Way, which has a diameter of 3 Mpc and contains about 60 galaxies, including both satellite galaxies and classical galaxies, such as M31\footnote{Andromeda Galaxy in the Messier catalog.} or M33\footnote{Triangulum Galaxy in the Messier catalog.} \citep{40_Bovill_2011, NASA/IPAC}.

\subsubsection{Milky Way}
The Milky Way is the SBc-type galaxy \citep{60_Gerhard_2002}, which houses the solar system, and therefore the Earth; which is located in the local arm or Orion arm. It is the second most massive galaxy in the local group, after M31. At its core we find a SMBH\footnote{From English, Super Massive Black Hole.}, from which thanks to \citet{3_Abuter_2019} we are approximately 8.1 Kpc from this galactic center; this location and distance from the galactic center becomes relevant when determining distances to other nearby galaxies using geometric methods, as well as in the use of galactic coordinates.

\subsubsection{Large Magellanic Cloud}
The Large Magellanic Cloud (LMC) is a satellite galaxy of the Milky Way (MW\footnote{From English, Milky Way.}) that extends for more than 7° in the southern celestial hemisphere. Its great extension was responsible for its discovery, its current name is given to it by the expedition led by Fernando Magellan in 1519 \citep{20Meschin_2011}.

\begin{figure*}[h!]
\centering
\caption{LMC spiral arms. This image represents a composite of two images: at the top left is the LMC in the optical band, while at the top right is the LMC in the hydrogen $\alpha$ line. At the bottom we find the stacking of the previous ones, where the image of the optical band is inverted for a correct stacking.}
\includegraphics[width=12cm]{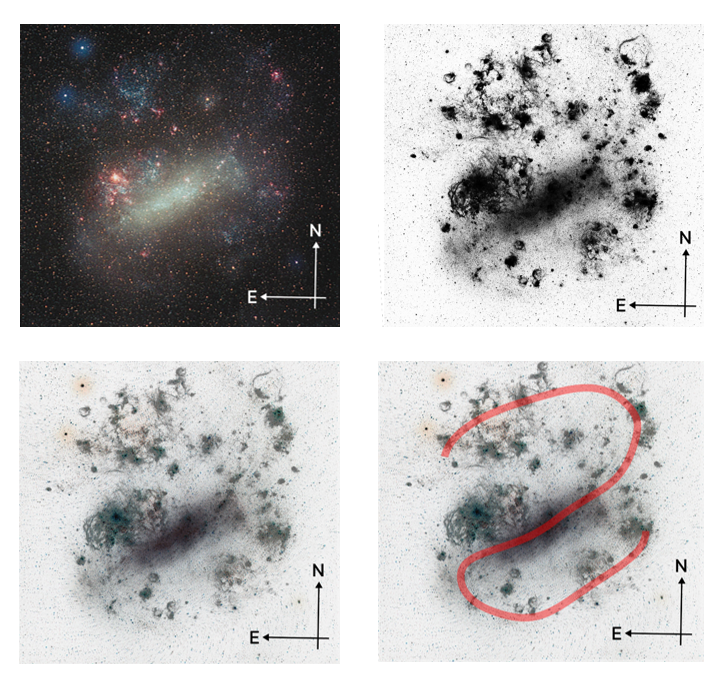}\\
\label{fig:LMC_Composite}
\end{figure*}

According to \citet{20Meschin_2011}, "the LMC is classified as a SB(s)m type galaxy, meaning that it is a barred spiral galaxy (SB) without a ring structure (s) of the Magellanic (m) type" (p. 3). However, the galaxy's barred spiral characteristic cannot be seen in the visible spectrum, only the central bulge can be seen; figure \ref{fig:LMC_Composite} shows a composition of images of the LMC in the visible (V) band thanks to \citet{LMC_ESA_OP} and in the hydrogen $\alpha$ band adapted from \citet{LMC_HA}, where the spiral arms can be seen.

As can be seen in figure \ref{fig:LMC_Composite}, the spiral arms do not appear to follow a symmetrical structure; this is because LMC has an inclination of $i = 25.110^{\circ} \pm0.365^{\circ}$ with respect to our plane \citep{59_Sukanta_2018}, which is essential for making corrections to the point where the distance is measured.

\subsubsection{Parallax}
Parallax is a geometric method for determining distances that has been used since 1838, when astronomer Friedrich Bessel published the first distance to a celestial body using parallax \citep{61_ESA_2019}. Since then, parallax has been used to determine distances to nearby celestial bodies, as it depends on the apparent displacement of the bodies and this displacement is reduced as soon as the body is too far from the observer, since the apparent motion will be minimal or null. ESA's GAIA mission made important contributions by making parallax measurements of more than 1.3 billion objects, mostly stars \citep{GAIA_1, GAIA_2, GAIA_6}.

\subsubsection{Intrinsic Variable Stars}
Variable stars are stars in which their apparent brightness observed from Earth presents changes, these are due to intrinsic or external factors of the star.

Due to the research approach, we will focus on intrinsic factors, one of which is pulsating stars, in turn there we find Cepheid variables, Cepheid type II variables and RR Lyrae type variables; each of which follows a different period-luminosity (PL) trend \citep{64_Soszynski_2018}.

 The period-luminosity relation was discovered by the American astronomer Henrietta Swan Leavitt in 1912, when she analyzed more than 1,700 stars from which she chose a group of 25 that are located in the Small Magellanic Cloud (SMC) and from there she determined this relation, which in turn allows us to obtain the absolute magnitude of each star and therefore its distance modulus $\mu$; this method is now used for the calibration of procedures and for the determination of distances with great precision from the local group onward \citep{23_Manahel_2013, Leavitt_1912, 45_Chen_2013, 59_Sukanta_2018}.

\subsubsection{Recession Velocity}
The recession velocity arises from the Hubble-Lemaître law, which relates the distance to an object to its recession velocity using its redshift and the Hubble parameter ($H_{0}$); all this in a universe that is constantly expanding and takes a value according to \citet{12_Riess_2023} of 67.4 $Km\: \, s^{-1}\: \,Mpc^{-1}$ being verified by the JWST.

\subsubsection{Eclipsing Binaries}
Eclipsing binaries are gravitationally interacting star systems whose orbits occasionally allow one of these stars to be hidden or eclipsed by the other from Earth; this causes changes in the apparent magnitude of each system \citep{62_BAA_2012}. Using eclipsing binaries as supported by the research of \citet{63_Graczyk_2021}, color index relationships can be established with respect to parameters such as surface brightness ($S_{v}$), which was essential in the procedure performed by \citet{5_Pietrzyński_2013}.

\begin{figure*}[h!]
\centering
\caption{Map of EBs captured by OGLE III. This figure is a composite of the visible spectrum image captured by \citet{LMC_ESA_OP} and the distribution of EBs as published by \citet{III_OGLE_ECL}. Each star according to its magnitude is represented in red, green or blue.}
\includegraphics[width=13cm]{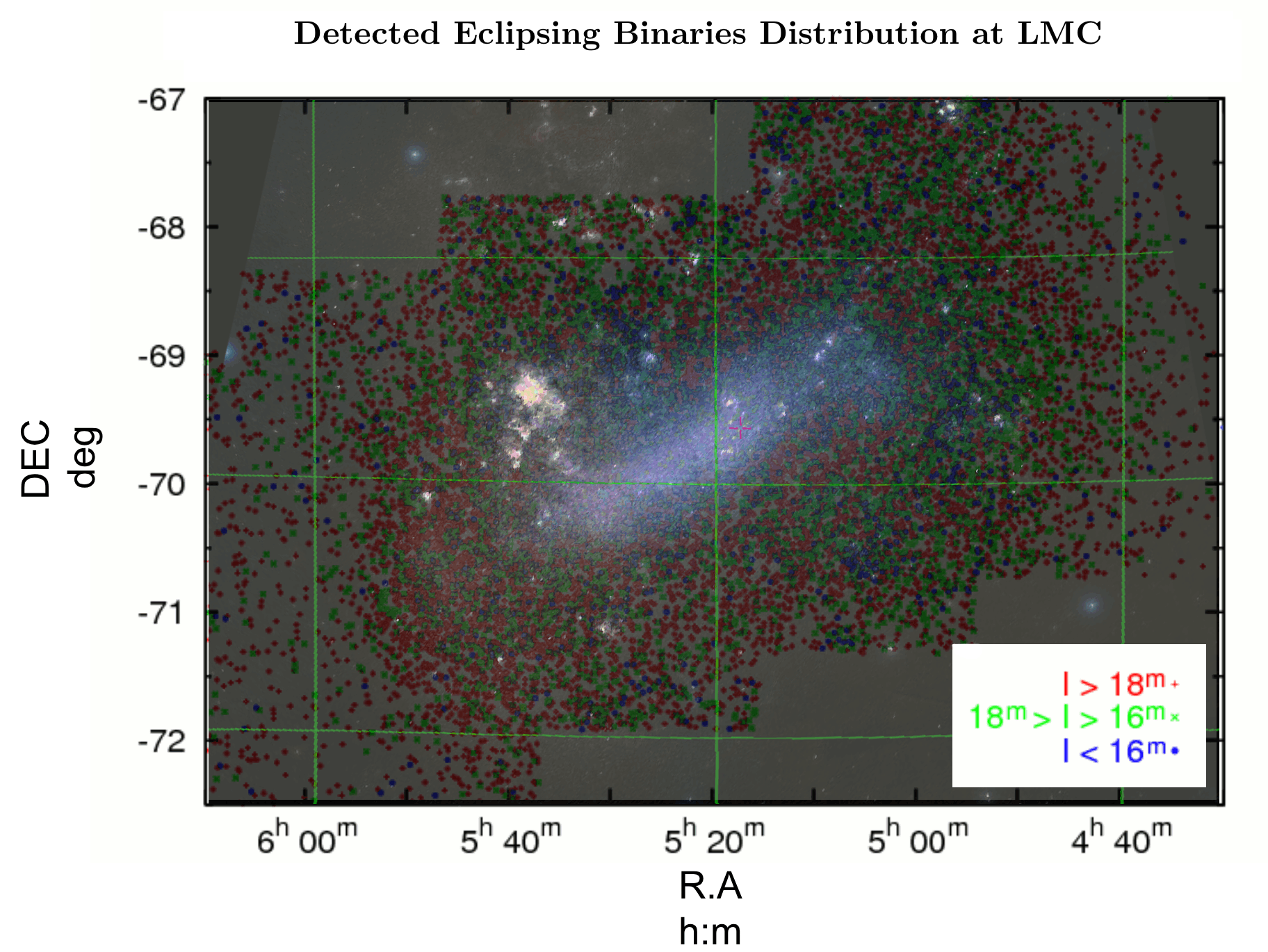}\\
\label{fig:ECL_LMC}
\end{figure*}

In the project presented by \citet{III_OGLE_ECL} called \textit{OGLE III}, which contains 26,121 eclipsing binary stars (EBs) in LMC, the distribution of stars found in the OGLE III EB catalogue can be seen in figure \ref{fig:ECL_LMC}.

Of the 26,121 star contributions, \citet{5_Pietrzyński_2013} took into account the systems\footnote{OGLE catalog systems following the nomenclature structure: OGLE-LMC-ECL-NNNNN.} 03160, 01866, 09660, 10567, 26122, 09114, 06575 and 15260; which are characterized by being intermediate-mass giant stars. In this way, the value of $d_{LMC}$ = $49.97$ Kpc was determined, with an estimated precision of 2.2 $\%$ \citep{5_Pietrzyński_2013}. This value is used as a reference in this research in order to compare the methods used.

\subsubsection{OGLE}
OGLE, for its acronym in English, is the Gravitational Optics Experiment; which began with the objective of searching for dark matter through the use of gravitational microlensing and has had great contributions and recognitions; the project currently has 4 phases (OGLE-I, OGLE-II, OGLE-III and OGLE IV). Its first phase began in 1992 and the experiment is still in progress. Most of the recorded observations have been made using the Las Campanas Observatory in Chile \citep{75_OGLEWEB}.

The stars of the LMC and SMC are of great interest to the OGLE project because they are the most affected by the gravitational lenses of the galactic halo; this is one of the main reasons why much of OGLE has focused on studying these regions.

\subsubsection{James Webb Space Telescope}
The James Webb Space Telescope is a 6.5-meter-aperture infrared telescope launched on 25 December 2021. Webb is one of the next-generation telescopes resulting from an international collaboration between NASA, ESA and the CSA \citep{76_WEBBGENERAL}. Webb's primary goal is to study every phase of the history of our universe.

This investigation will use measurements made with Webb's instruments: MIRI and NIRSpec.

\begin{itemize}
    \item MIRI: The Mid-Infrared Instrument features a camera and spectrograph focused on analyzing the mid-infrared part of the electromagnetic spectrum, covering the wavelength range of 5 to 28 $\mu m$; MIRI achieves this sensitivity by maintaining an operating temperature of 7 $^{\circ}$K \citep{78_WEBBMIRI}.
    \item NIRSpec: The Near Infrared Spectrograph operates in the range of 0.6 to 5 $\mu m$ and is capable of making 100 simultaneous measurements thanks to innovations in its construction \citep{77_WEBBNIRS}.
\end{itemize}

\subsubsection{Units of Distance in the Universe}
When measuring distances in the universe we find different units that are commonly used despite not being directly in the international system. These units arise as a response to the problem of scales in the universe; when measuring distances in the solar system we use the international system and the astronomical units (AU)\footnote{From English, Astronomical Unit.}, when measuring distances in the surroundings of the solar system we use light years, when talking about distances beyond the solar system, pc\footnote{Distance unit, parsec.} and Mpc \citep{11_NASA_Distances_2020} are used. The aforementioned units are represented in the table \ref{tab:distance_units} with their respective equivalence in the international system.

\begin{deluxetable}{lll}
\tablewidth{0pt}
\tablecaption{Cosmic Distance Units \label{tab:distance_units}}
\tablehead{
\colhead{Name} & \colhead{Symbol} & \colhead{Equivalence in the SI\footnote{Acronyms, International System.}}
}
\startdata
Astronomical Unit & AU & $1.4959\times10^{11}m$ \\
Light Year & ly & $2.4605\times10^{15}m$ \\
Parsec & pc & $3.0856\times10^{16}m$ \\
\enddata
\tablecomments{This table provides conversions between common astronomical distance units and their equivalent values in the International System of Units.}
\end{deluxetable}

\subsubsection{Number Formatting}
It is clarified that for this research the International System of Units is used, in this case, the point denotes decimal separation ($1.5 = 3/2$). Likewise, during the research, up to 4 decimal places are used whenever possible and following the recommendations of \citet{9_Chen_2022} each value obtained has its respective systematic error and standard error.

\section{Methodology} \label{sec:methodology}

\subsection{Type and Scope of Study}
This is a work of approach to methodology and research which is carried out as a Bachelor's degree project at the Colegio Andino de Tunja. During the development of the project, tutoring and advice from teachers from the Colegio Andino de Tunja was provided.

\subsection{Sources of Information}
Most of the information taken as reference comes from scientific journals: \href{https://iopscience.iop.org/journal/0004-637X}{\textit{The Astrophysical Journal}}, \href{https://www.aanda.org/}{\textit{Astronomy and Astrophysics}}, \href{https://ui.adsabs.harvard.edu/}{\textit{astrophy sics data system}} and \href{https://arxiv.org/archive/astro-ph}{\textit{arxiv - Astrophysics}}. Likewise, the following databases were mostly used: \href{https://simbad.unistra.fr/simbad/}{SIMBAD} and \href{https://ned.ipac.caltech.edu/}{NASA/IPAC Extragalactic Database}.

\subsection{Data Analyzed}
In this study, the following data have been analyzed by the distance calculation method: parallax, parallax angle ($\pi$); variable stars, light curves, PL ratios, distance modules ($\mu$), interstellar extinction; eclipsing binaries, angular size ($mas$), distance module ($\mu$); recession velocity, Hubble parameter ($H_{0}$), redshift ($z$); Kepler's third law, orbital period, semi-major axis, mass of the Milky Way ($M_{MW}$); supernova SN1987a, angular size, radius.

\subsection{Sources of Information}
The analyzed data were retrieved from \href{https://archive.stsci.edu/}{\textit{Mikulski Archive for Space Telescopes}} (MAST) in the case of the analyzed data coming from the JWST, HST and IUE space telescopes\footnote{From English, International Ultraviolet Explorer.}. Likewise, the GAIA space telescope was used through the \textit{Python} library: \href{https://astroquery.readthedocs.io/en/latest/gaia/gaia.html}{astroquery.gaia}. Finally, the catalog of eclipsing binaries in LMC open to the public thanks to the collaboration of \citet{OGLE_Wyrzykowski_2003} in the OGLE project was used.

\subsection{Methodological Design}
In order to calculate the distance to LMC, a statistical analysis has been carried out using 6 different methods for determining this distance, as well as the comparison of each method; which are described in the following sections.

\subsubsection{Analysis Variables}
The distance to the LMC defined in the document has been estimated as ($d_{LMC}$). Similarly, the variables previously presented are included in this research.

In determining this distance, the inclination provided by the research of \citet{59_Sukanta_2018} of $i = 25.110^{\circ} \pm0.365^{\circ}$ in the geometric methods of determining distances is taken into account.

\subsubsection{Programs Used}
Scripts made in the Python programming language 3.9.13 have been used to obtain information from GAIA, as well as its subsequent analysis; the creation of the figure \ref{fig:LocalGroup}. Likewise, use was made of the tools and utilities incorporated in MAST; the platform \href{https://js9.si.edu/}{JS9} for processing FITS images\footnote{From English, Flexible Image Transport System.} and \textit{Microsoft Excel} for the creation of graphs and statistical analysis of the procedures.

\section{Procedure} \label{sec:procedure}
The procedure used to determine the distance to the Large Magellanic Cloud using 6 different methods is presented below.

\subsection{Parallax}
Using the GAIA data and integrating it into Python, a script is created\footnote{See \hyperlink{page.20}{appendix} B.} which identifies 696,321 stars \citep{GAIA_1, GAIA_2, GAIA_3, GAIA_4, GAIA_5, GAIA_6, GAIA_7} which are located within a 12º radius around the absolute equatorial coordinate of the LMC center ($\alpha, \delta$)$_{J2000}$ ($5^{h}17^{m}$, $-69^{\circ}{02}'$).\citep{20Meschin_2011}. Subsequently, of these 696,321\footnote{For the total record of stars used see project repository} stars, those without parallax information were discarded, thus obtaining 576,121 stars; from which the parameters indicated in table \ref{tab:Parallax_DA} are analyzed, in which the sample of stars is cut to 336,975; establishing a tolerance range of $0.5\: \,\sigma$.

\begin{deluxetable}{llll}[h!]
\tablewidth{0pt}
\tablecaption{Data Analysis: Parallax \label{tab:Parallax_DA}}
\tablehead{
\colhead{Stdv.} & \colhead{Avg. Parallax} & \colhead{$+ 0.5\: \,\sigma$} & \colhead{$- 0.5\: \,\sigma$} \\
\colhead{$\sigma$} & \colhead{$mas$} & \colhead{$mas$} & \colhead{$mas$}
}
\startdata
0.9858 & 0.2576 & 0.7506 & -0.2352 \\
\enddata
\tablecomments{The following accepted range of $\pi$ is established: $[-0.2352 \, \text{mas},\: 0.7506 \, \text{mas}]$.}
\end{deluxetable}

Finally, transforming $\pi$ from $mas$ to $arcsec$, the approximation when $\theta << 1$ of $tan(\theta)\: \,\approx\: \,\theta$ is used, thus obtaining $d_{LMC}$ in parsecs by the equation \ref{eq:parallax}.
\begin{equation}
d_{LMC} = \frac{1}{\pi_{(arcsec)}}
\label{eq:parallax}
\end{equation}
        
The distance in Kpc to each of the 336,975 stars is determined. In addition, its average, standard deviation ($\sigma$) and standard error are obtained. Similarly, to obtain the systematic error it was taken into account that GAIA has the following uncertainties: for stars with magnitudes greater than 14 $\pm\: \,0.04 mas$, for stars with magnitudes between 14 and 17 $\pm\: \,0.1 mas$ and for stars with magnitudes less than 17 $\pm\: \,0.7 mas$ \citep{65_LuriGAIAERROR_2018}. Thus, the data mentioned in table \ref{tab:Parallax_R} are presented.

\begin{deluxetable}{ccc}[h!]
\tablewidth{0pt}
\tablecaption{Distance to LMC using parallax \label{tab:Parallax_R}}
\tablehead{
\colhead{\textbf{$d_{LMC}$}} & \colhead{Standard Error} & \colhead{Systematic Error} \\
\colhead{Kpc} & \colhead{Kpc} & \colhead{Kpc}
}
\startdata
41.2459 & $\pm$ 3.5611 & $\pm$ 0.3991 \\
\enddata
\end{deluxetable}

\begin{deluxetable*}{ccccccc}

\tablewidth{0pt}
\tablecaption{Data collected from GAIA \label{tab:GAIA_Parallax}}
\tablehead{
\colhead{ID$^{*}$} & \colhead{$\alpha_{J2000}$} & \colhead{$\delta_{J2000}$} & \colhead{$\pi$} & \colhead{Error} & \colhead{G} & \colhead{$d_{LMC}$} \\
\colhead{} & \colhead{deg} & \colhead{deg} & \colhead{mas} & \colhead{mas} & \colhead{mag} & \colhead{Kpc}
}
\startdata
465454800533777 & 73.1239 & -71.7573 & 0.0186 & $\pm$ 0.7 & 18.96 & 53.7380 \\
465453361291771 & 74.5500 & -71.5012 & 0.0385 & $\pm$ 0.1 & 15.54 & 25.9904 \\
466055477912470 & 80.4833 & -65.8301 & -0.0166 & $\pm$ 0.04 & 11.62 & 60.3212 \\
\enddata
\tablenotetext{*}{ID refers to the Gaia source identifier.}
\tablenotetext{a}{Apparent magnitude (mag) in the G band.}
\tablenotetext{b}{From English, degrees.}
\tablecomments{Table \ref{tab:GAIA_Parallax} shows us 3 of the 336,975 records obtained by GAIA and the format in which they were used as can be found in the repository of this research.}
\end{deluxetable*}

\subsection{Lyrae RR type variables}
In the present investigation, RR Lyrae type stars will be used to determine the distance to the LMC by using the distance modulus ($\mu$) and the PL relation. Thanks to \citet{20Meschin_2011} we know that there are more than 24,900 RR Lyrae type variable stars in the LMC.

\begin{figure}
    \centering
    \includegraphics[width=8.5cm]{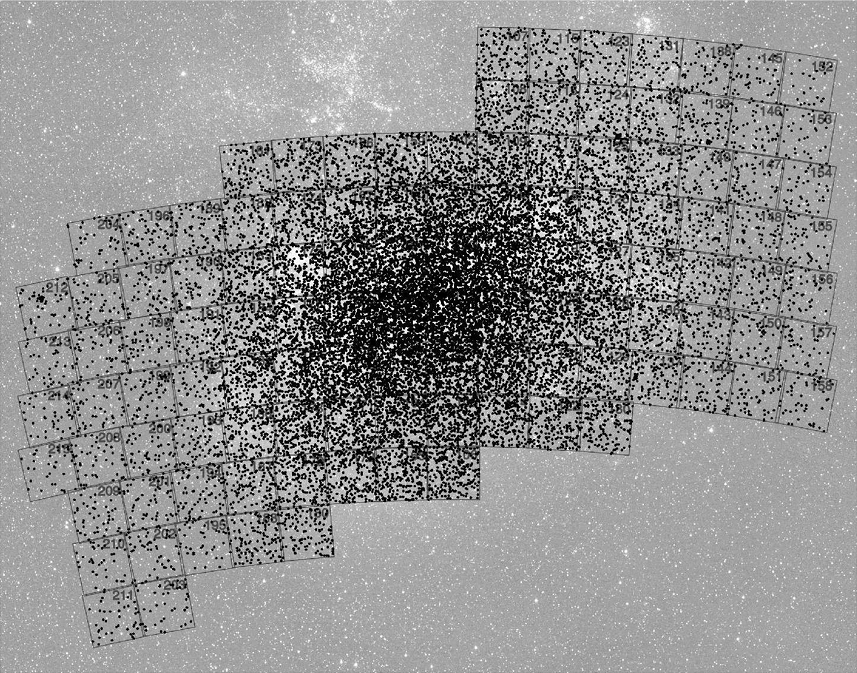}
    \caption{Map of the Lyrae RRs captured by OGLE III. The figure \ref{fig:distanceschrono} represents the location of stars recorded by OGLE III \citep{III_OGLE_RRLyrae}.}
    \label{fig:RR_LYRAE_LMC}
\end{figure}

In the case of RR Lyrae type variables, the publicly available OGLE III catalogue was used\footnote{From English, Optical Gravitational Lensing Experiment.} \citet{III_OGLE_RRLyrae} and 17,693 stars of the RR Lyrae type were selected. The location of these can be seen in figure \ref{fig:RR_LYRAE_LMC} \citep{III_OGLE_RRLyrae}.

 Considering that we have variable stars of the RR Lyrae type, they present a PL relation in the infrared band. The PL relation as presented by \citet{67_Catelan_2004} in this case is given by:
 
\begin{equation}
M_{I} = 0.471 - 1.132log(P_{days}) + 0.205log(z)
\label{eq:PL_RRLYRAE}
\end{equation}
\textit{Note.} In the equation \ref{eq:PL_RRLYRAE} we have $P$ which represents the period of the star in days and $z$, which is a metallicity coefficient.

As we can see in the equation \ref{eq:PL_RRLYRAE} we need the value $z$. Also, in the figure \ref{fig:RR_LYRAE_LMC} we can see that the majority of the star samples are located in the galactic bulge, thanks to this location and the research of \citet{66_Choudhury_2021} we have that $z = 0.008\: \, \pm 0.001$. Applying the PL relation and the now known value of $z$, we can find the absolute magnitude ($M_{I}$) to in turn find the distance modulus ($\mu$) and then determine the distance by:

\begin{equation}
\mu = m - M
\label{eq:DisModulus}
\end{equation}

\begin{equation}
d_{LMC} = 10 ^ {(\frac{\mu + 5}{5})}
\label{eq:Distancemu}
\end{equation}

\begin{deluxetable*}{ccccccc}

\tablewidth{0pt}
\tablecaption{RR Lyrae variable data collected from OGLE III \label{tab:OGLE_RRLYRAE_DATA}}
\tablehead{
\colhead{ID$^{*}$} & \colhead{$m_{v}$} & \colhead{$m_{I}$} & \colhead{$P$} & \colhead{Error (P)} & \colhead{$M_{I}$} & \colhead{$d_{LMC}$} \\
\colhead{} & \colhead{mag} & \colhead{mag} & \colhead{days} & \colhead{days} & \colhead{mag} & \colhead{Kpc}
}
\startdata
OGLE-LMC-RRLYR-09258 & 13.23 & 13.77 & 0.5534047 & $\pm$ 0.0000006 & 0.34 & 3.7800 \\
OGLE-LMC-RRLYR-19082 & 18.46 & 19.04 & 0.6064705 & $\pm$ 0.0000002 & 0.29 & 43.0629 \\
OGLE-LMC-RRLYR-11840 & 18.63 & 19.16 & 0.6247962 & $\pm$ 0.0000013 & 0.28 & 46.7123 \\
\enddata
\tablenotetext{*}{ID refers to the OGLE source identifier.}
\tablecomments{Table \ref{tab:OGLE_RRLYRAE_DATA} shows 3 of the 17,693 records obtained by OGLE and the format in which they were used, as can be found in the repository of this research.}
\end{deluxetable*}

\begin{deluxetable*}{cccccccccccc}

\tablewidth{0pt}
\tablecaption{Cepheid data collected from OGLE III \label{tab:OGLE_CEPHEIDS_DATA}}
\tablehead{
\colhead{ID} &
\colhead{$m_{I}$} & \colhead{$m_{V}$} & \colhead{Period} & \colhead{Uncertainty (P)} & \colhead{$M_{I}$} & \colhead{$M_{V}$} & \colhead{$\mu_{W}$} & \colhead{$d_{LMC}$ ($V$)} & \colhead{$d_{LMC}$ ($I$)} & \colhead{$d_{LMC}$ ($\mu_{W}$)} \\
\colhead{} & \colhead{mag} & \colhead{mag} & \colhead{days} & \colhead{days} & \colhead{mag} & \colhead{mag} & \colhead{mag} & \colhead{Kpc} & \colhead{Kpc} & \colhead{Kpc}
}
\startdata
0986 & 12.41 & 13.35 & 31.0504 & $\pm$ 0.000482 & -4.96 & -6.03 & 18.50 & 45.8526 & 48.6819 & 51.1498 \\
1058 & 12.48 & 13.39 & 30.3989 & $\pm$ 0.000517 & -4.93 & -6.00 & 18.54 & 46.1154 & 49.5990 & 48.9571 \\
0821 & 12.61 & 13.75 & 25.8032 & $\pm$ 0.000113 & -4.74 & -5.79 & 18.49 & 49.9843 & 47.8635 & 50.5916 \\
0535 & 12.64 & 13.57 & 17.2755 & $\pm$ 0.000149 & -4.28 & -5.27 & 17.45 & 37.1183 & 38.1814 & 50.5667 \\
2019 & 12.70 & 13.64 & 28.1034 & $\pm$ 0.000237 & -4.84 & -5.90 & 18.53 & 49.6177 & 52.5932 & 48.3640 \\
\enddata
\tablenotetext{*}{The identifier prefix for each star has been omitted, leaving only its numerical value. Prefix: OGLE-LMC-CEP-.}
\tablecomments{Five of the 1,804 stars are shown in Table \ref{tab:OGLE_CEPHEIDS_DATA}.}
\end{deluxetable*}

\begin{deluxetable}{ccccc}
\tablewidth{0pt}
\tablecaption{Distance by Classical Cepheid stars \label{tab:CEPHEIDS_R}}
\tablehead{
\colhead{Variables} & \colhead{$d_{LMC}$ (V)} & \colhead{$d_{LMC}$ (I)} & \colhead{$d_{LMC}$ ($\mu$)} & \colhead{$d_{LMC}$} \\
\colhead{} & \colhead{Kpc} & \colhead{Kpc} & \colhead{Kpc} & \colhead{Kpc}
}
\startdata
Avg. & 50.4981 & 50.2015 & 50.7794 & 50.4930 \\
Std. Error & $\pm$ 0.2173 & $\pm$ 0.1312 & $\pm$ 0.0747 & $\pm$ 0.1411 \\
Sys. Error & $\pm$ 0.0001 & $\pm$ 0.0002 & $\pm$ 0.0003 & $\pm$ 0.0002 \\
\enddata
\tablecomments{The distance to the Large Magellanic Cloud (LMC) was determined using Classical Cepheid stars in different bands. The total estimated distance is shown in the last column.}
\end{deluxetable}

Finally, in this way and by averaging the distances obtained with the 17,693 stars we obtain the distance to LMC using variable stars of type RR Lyrae, together with their respective standard and systematic errors as can be seen in table \ref{tab:RRLYRAE_R}.

\begin{deluxetable}{ccc}
\tablewidth{0pt}
\tablecaption{Distance via RR Lyrae stars \label{tab:RRLYRAE_R}}
\tablehead{
\colhead{$d_{LMC}$} & \colhead{Standard Error} & \colhead{Systematic Error} \\
\colhead{Kpc} & \colhead{Kpc} & \colhead{Kpc}
}
\startdata
49.9979 & $\pm$ 0.0503 & $\pm$ 0.0481 \\
\enddata
\tablecomments{The distance to the Large Magellanic Cloud (LMC) was estimated using RR Lyrae stars, including statistical and systematic uncertainties.}
\end{deluxetable}

\subsection{Classical Cepheid Variables}
\begin{figure}[h!]
\centering
\caption{Map of classical Cepheids captured by OGLE III. The figure shows the location of the selected stars in yellow \citep{III_OGLE_CEPHEIDS}}
\includegraphics[width=8.5cm]{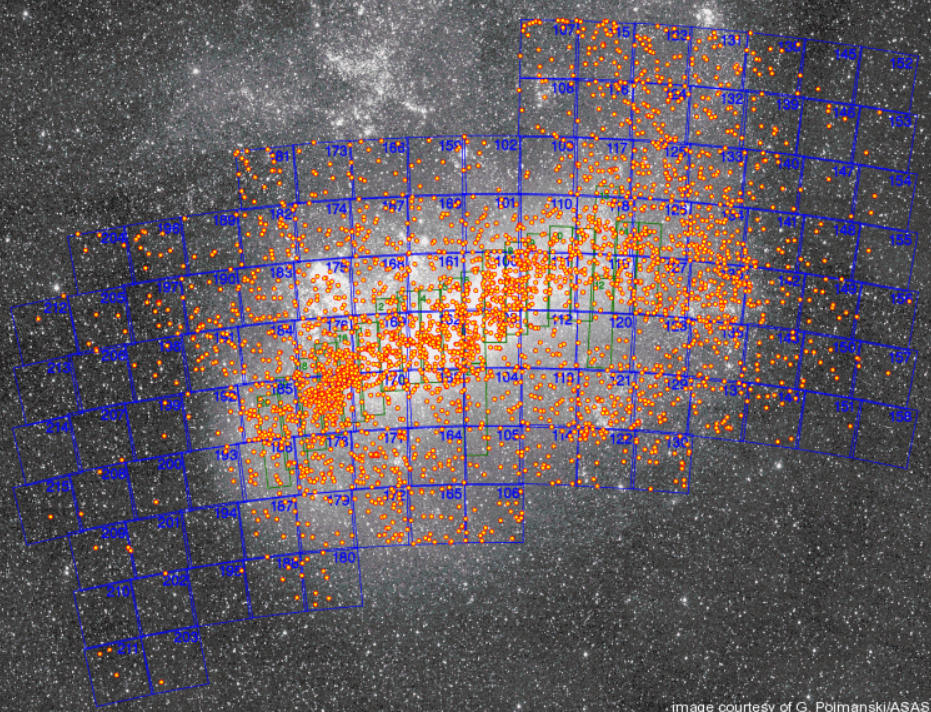}
\label{fig:CEPHEIDS_LMC}
\end{figure}

In the case of classical Cepheid variables, the OGLE catalogue was used in the same way, in this case for Cepheid variables \citep{III_OGLE_CEPHEIDS} where records of 1.80e stars were selected, the location of these can be found in figure \ref{fig:CEPHEIDS_LMC}.

The following PL relationships are used according to the band being analyzed: $M_{V}$ and $M_{I}$ adapted from the research of \citet{72_Madore_2017}; and the distance modulus directly $\mu_{W}$ adapted from the research of \citet{71_Ngeow_2012} using the color index $(VI)$ and the Wesenheit function. In the case of $M_{V}$ and $M_{I}$, the interstellar extinctions were adjusted\footnote{Correction made taking into account the dust masses that can interfere with the results.} presented by \citet{69_Imara_2007} with a value of $A_{V}\: \,=\: \,0.3\: \,mag$, and with the contributions of \citet{70_USM_1999} we have that $A_{I}\: \,=\: \,0.48\times A_{v}\: \,mag$; in the case of $\mu_{W}$, the interstellar extinction is not taken into account since internally the equation takes into account the color index $(VI)$ which takes into account the extinction. In this way we obtain the equations \ref{eq:CEPH_1_V}, \ref{eq:CEPH_2_I} and \ref{eq:CEPH_3_mu}.
    
\begin{equation}
M_{V} = - 2.670[log(P_{days})-1.0] - 3.944 + A_{V}
\label{eq:CEPH_1_V}
\end{equation}
    
\begin{equation}
M_{I} = - 2.983[log(P_{days})-1.0] - 4.706 + A_{I}
\label{eq:CEPH_2_I}
\end{equation}
    
\begin{equation}
\mu_{W} = I -1.55(V - I) + 3.313log(P_{days}) + 2.639
\label{eq:CEPH_3_mu}
\end{equation}

In table \ref{tab:OGLE_CEPHEIDS_DATA} you can see the data that were taken into account as well as the format in which they are found in the research repository.

Thus, with the distance module described in the equation \ref{eq:DisModulus} and the $d_{LMC}$ described in the equation \ref{eq:Distancemu} we determine the distance to the LMC using the visible, infrared bands and the color index (VI) described in the equation \ref{eq:CEPH_3_mu}. With the availability of these 3 distances, an average is made comparing their respective standard error and systematic error; data represented in table \ref{tab:CEPHEIDS_R}.

\subsection{Redshift}
When determining the distance to a body using the redshift, the provisions of the Hubble-Lemaître Law are taken into account, of which the equations \ref{eq:Hubble_1} and \ref{eq:Hubble_2} will be used. To do this, the parameter $z$ must be established, which represents the red or blue shift, defined in the equation \ref{eq:z}.
\begin{equation}
\nu = H_{0} \times d
\label{eq:Hubble_1}
\end{equation}
\begin{equation}
\nu = cz
\label{eq:Hubble_2}
\end{equation}
\begin{equation}
z = \frac{\lambda _{obs} - \lambda _{rest}}{\lambda _{rest}}
\label{eq:z}
\end{equation}

\begin{figure}[h!]
\centering
\caption{Map of LMC samples taken by JWST. The figure represents the location of the selected regions with a red dot. The image was created with the help of \citet{MAST} and its \textit{Astroview} tool.}
\includegraphics[width=8.5cm]{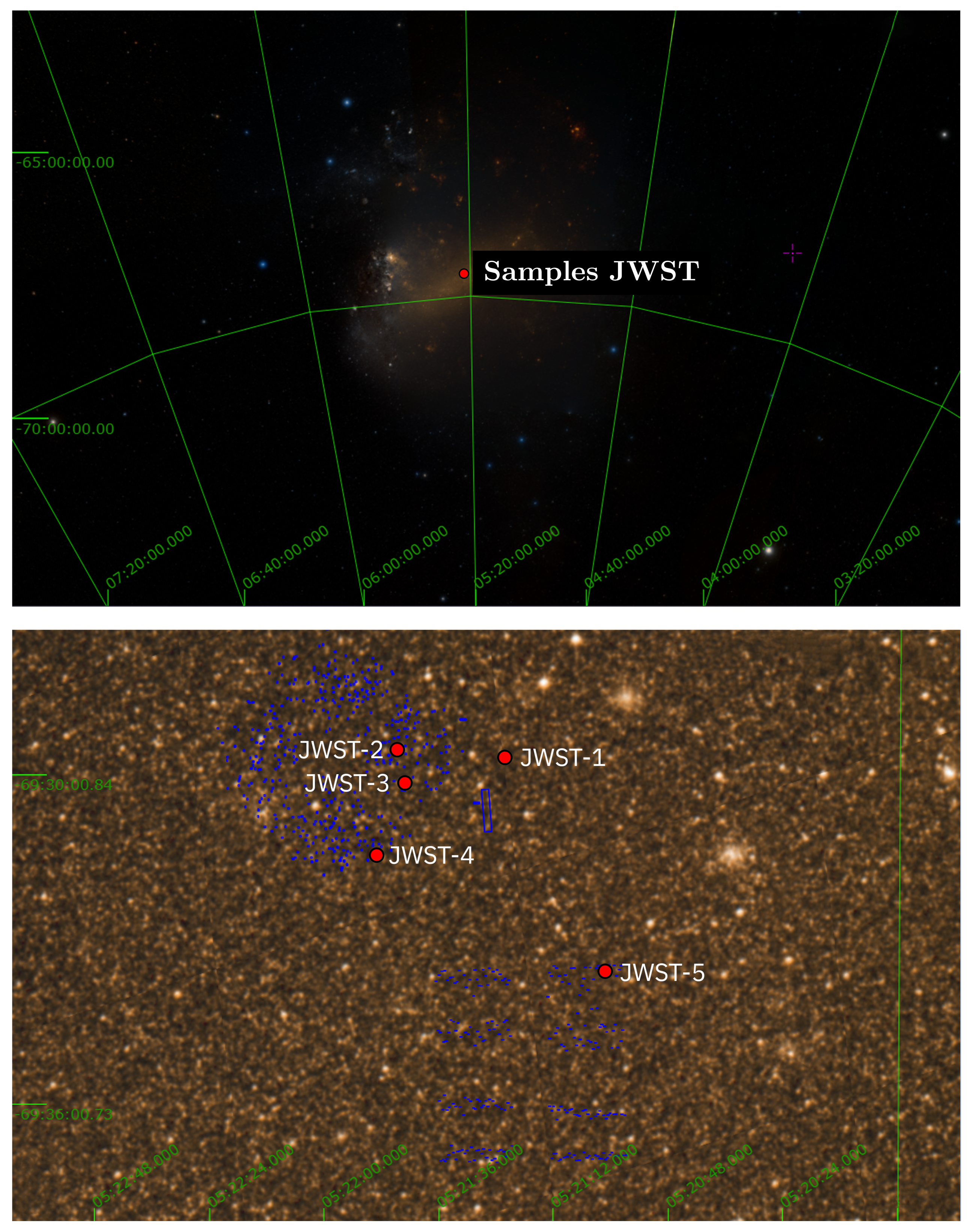}
\label{fig:JWST_LMC}
\end{figure}

Five regions of interest in LMC are selected below in order to determine the redshift by analyzing wavelengths of interest and their respective displacement. The JWST space telescope with the NIRSPEC/MSA and MIRI instruments was used, which allow observing the spectrum of these 5 regions. The location of the 5 analyzed regions is represented in figure \ref{fig:JWST_LMC} \citep{73_Sánchez_2023}. Additionally, these regions are detailed precisely in table \ref{tab:JWST_LMC_DET}.

The original spectrum of each region is then presented, accompanied by the spectrum with the adjustment of spectral lines in order to determine the redshift.

\begin{deluxetable*}{cccccccc}

\tablewidth{0pt}
\tablecaption{Regions selected in LMC using the JWST \label{tab:JWST_LMC_DET}}
\tablehead{
\colhead{ID$^{\ast}$} & \colhead{ID$^{\dagger}$} & \colhead{$\alpha_{J2000}$} & \colhead{$\delta_{J2000}$} & \colhead{Instrument} & \colhead{$\lambda_{min}$} & \colhead{$\lambda_{max}$} & \colhead{$t_{exp}$} \\
\colhead{} & \colhead{} & \colhead{deg} & \colhead{deg} & \colhead{} & \colhead{$\AA$} & \colhead{$\AA$} & \colhead{s}
}
\startdata
JWST-1 & jw01029-o009\_t007\_miri\_p750l & 80.3500 & -69.4942 & MIRI/SLIT & 30000 & 140000 & 299.704 \\
JWST-2 & jw01117-o007\_s00051\_nirspec\_clear-prism & 80.4923 & -69.4971 & NIRSPEC/MSA & 8500 & 53600 & 875.334 \\
JWST-3 & jw01117-o022\_s89012\_nirspec\_f100lp-g140m & 80.4913 & -69.4976 & NIRSPEC & 10000 & 19000 & 1313 \\
JWST-4 & jw01117-o032\_s34975\_nirspec\_f100lp-g140m & 80.4894 & -69.4964 & NIRSPEC & 9660 & 19000 & 2626 \\
JWST-5 & jw01473-o016\_s00048\_nirspec\_clear-prism & 80.4888 & -69.4994 & NIRSPEC/MSA & 5690 & 53400 & 1313 \\
\enddata
\tablecomments{ID$^{\ast}$ created for the present investigation; ID$^{\dagger}$ original ID from the MAST database; $t_{exp}$ is the capture exposure time. Table \ref{tab:JWST_LMC_DET} shows the 5 previously mentioned regions accompanied by the characteristics of each measurement.}
\end{deluxetable*}

\begin{figure*}
\centering
\caption{JWST-1 redshift results. This figure is a composition of the steps that were carried out to determine the $z$ of each region.}
\includegraphics[width=5.25in]{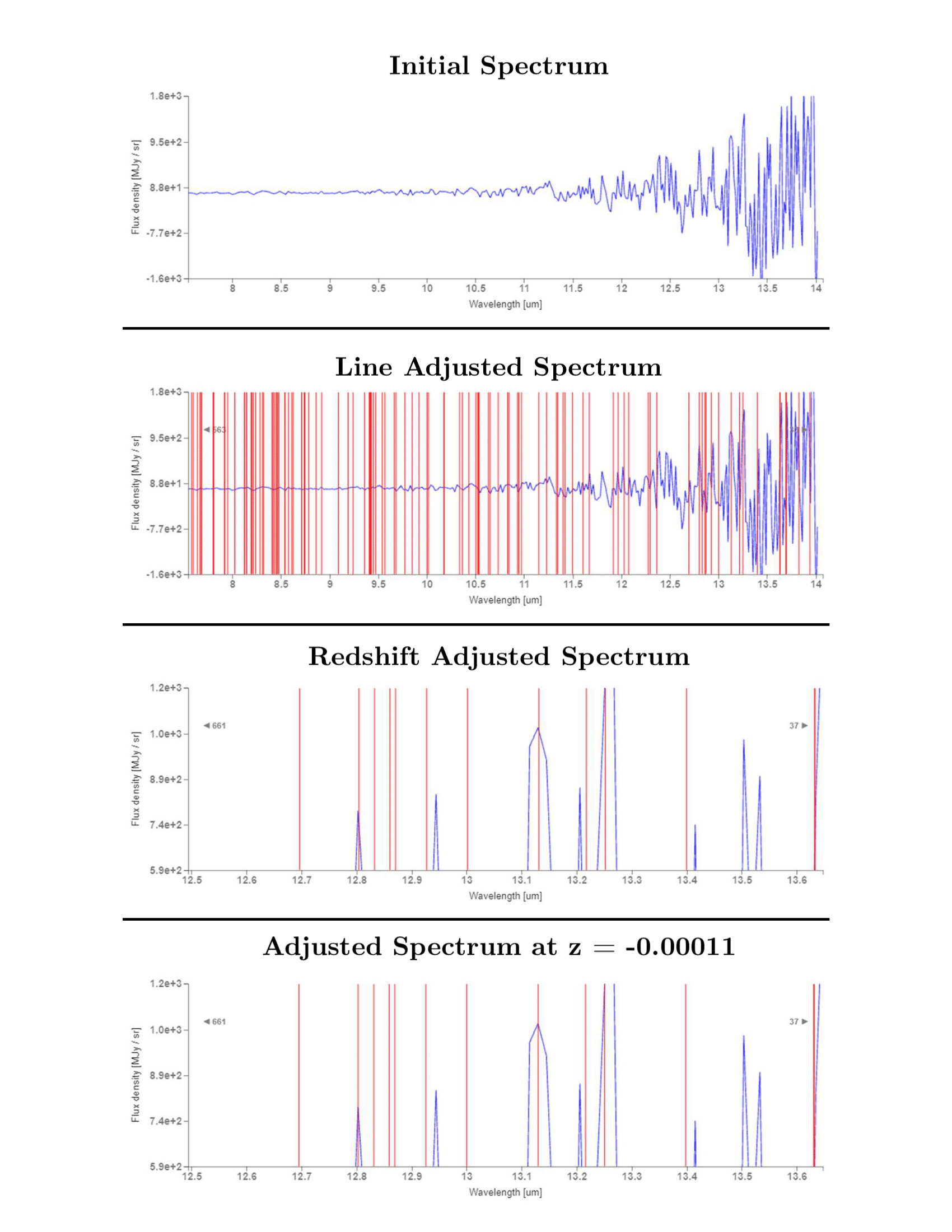}
\label{fig:JWST_1}
\end{figure*}

\begin{figure*}
\centering
\caption{JWST-2 redshift results. This figure is a composition of the steps that were carried out to determine the $z$ of each region.}
\includegraphics[width=5.25in]{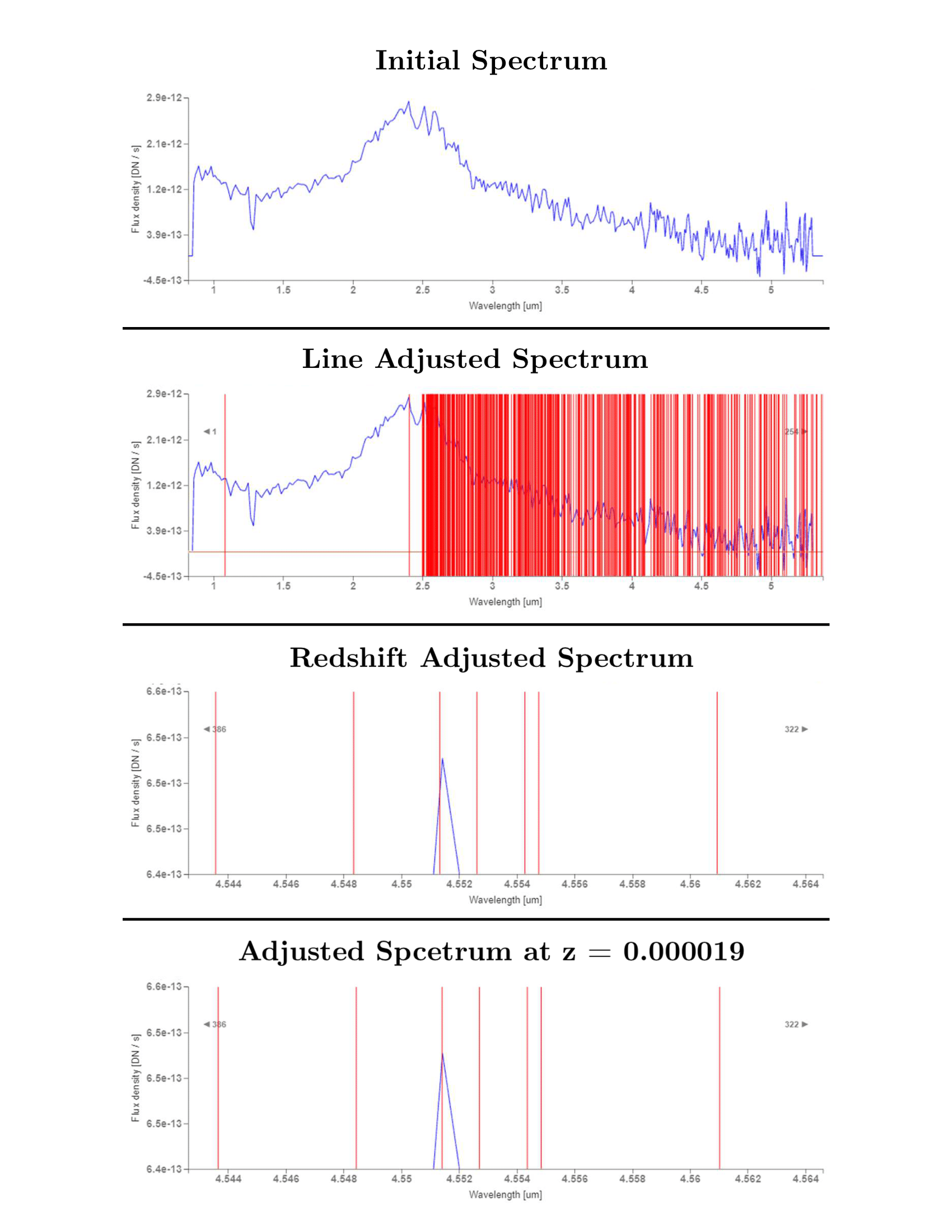}
\label{fig:JWST_2}
\end{figure*}

\begin{figure*}
\centering
\caption{JWST-3 redshift results. Figure \ref{fig:JWST_3} is a composition of the steps that were carried out to determine the $z$ of each region.}
\includegraphics[width=5.25in]{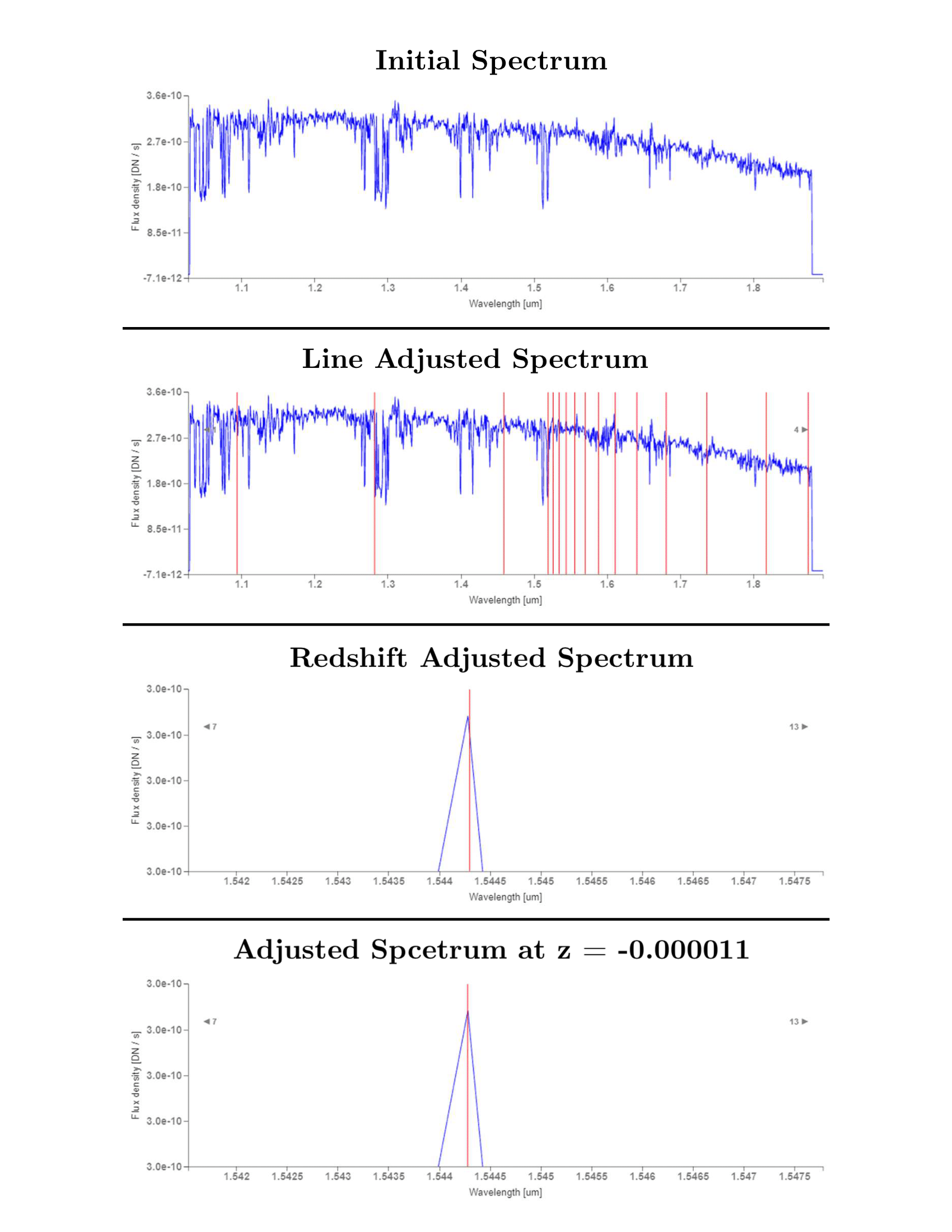}\\
\label{fig:JWST_3}
\end{figure*}

\begin{figure*}
\centering
\caption{JWST-4 redshift results. Figure \ref{fig:JWST_4} is a composition of the steps that were carried out to determine the $z$ of each region.}
\includegraphics[width=5.25in]{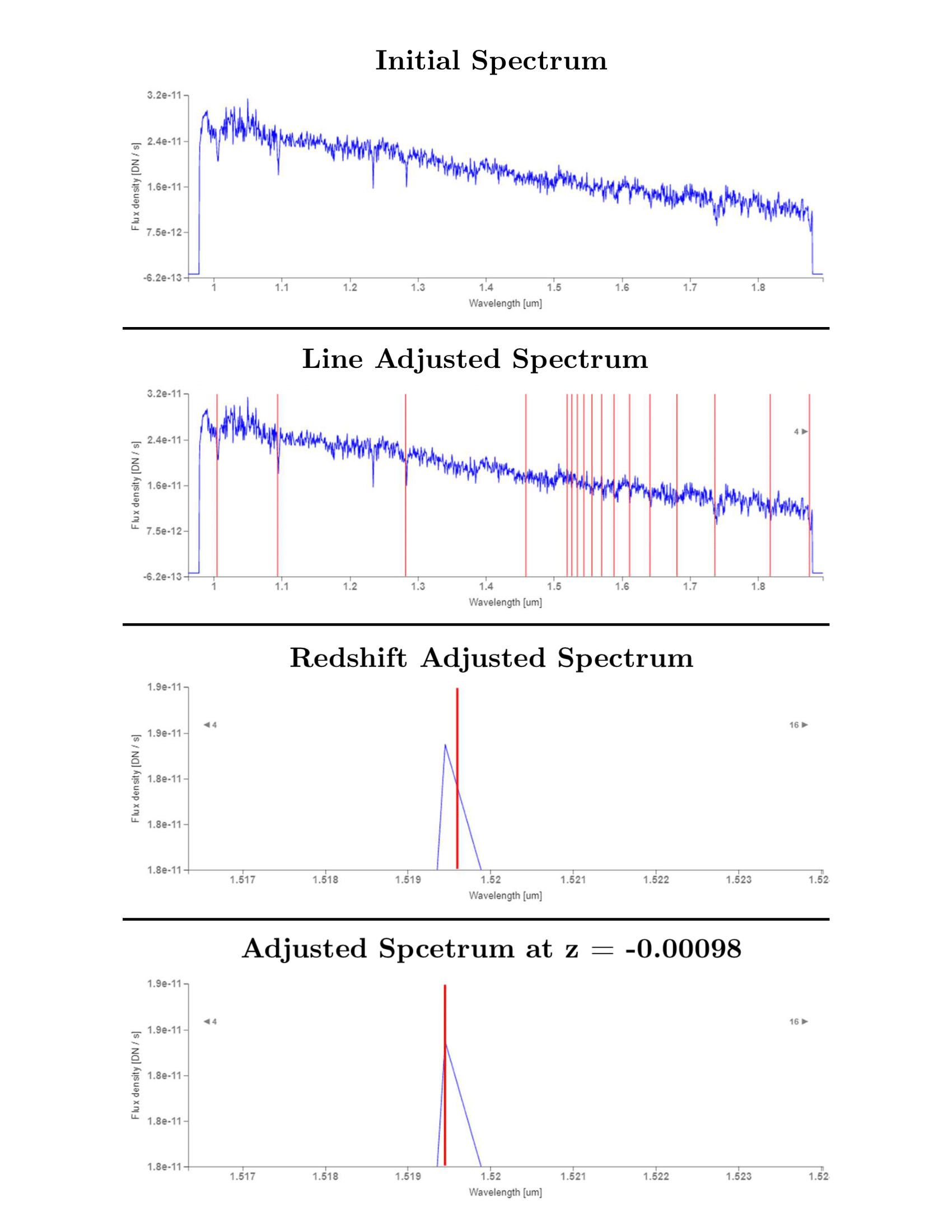}\\
\label{fig:JWST_4}
\end{figure*}

\begin{figure*}
\centering
\caption{JWST-5 redshift results.Figure \ref{fig:JWST_5} is a composition of the steps that were carried out to determine the $z$ of each region.}
\includegraphics[width=5.25in]{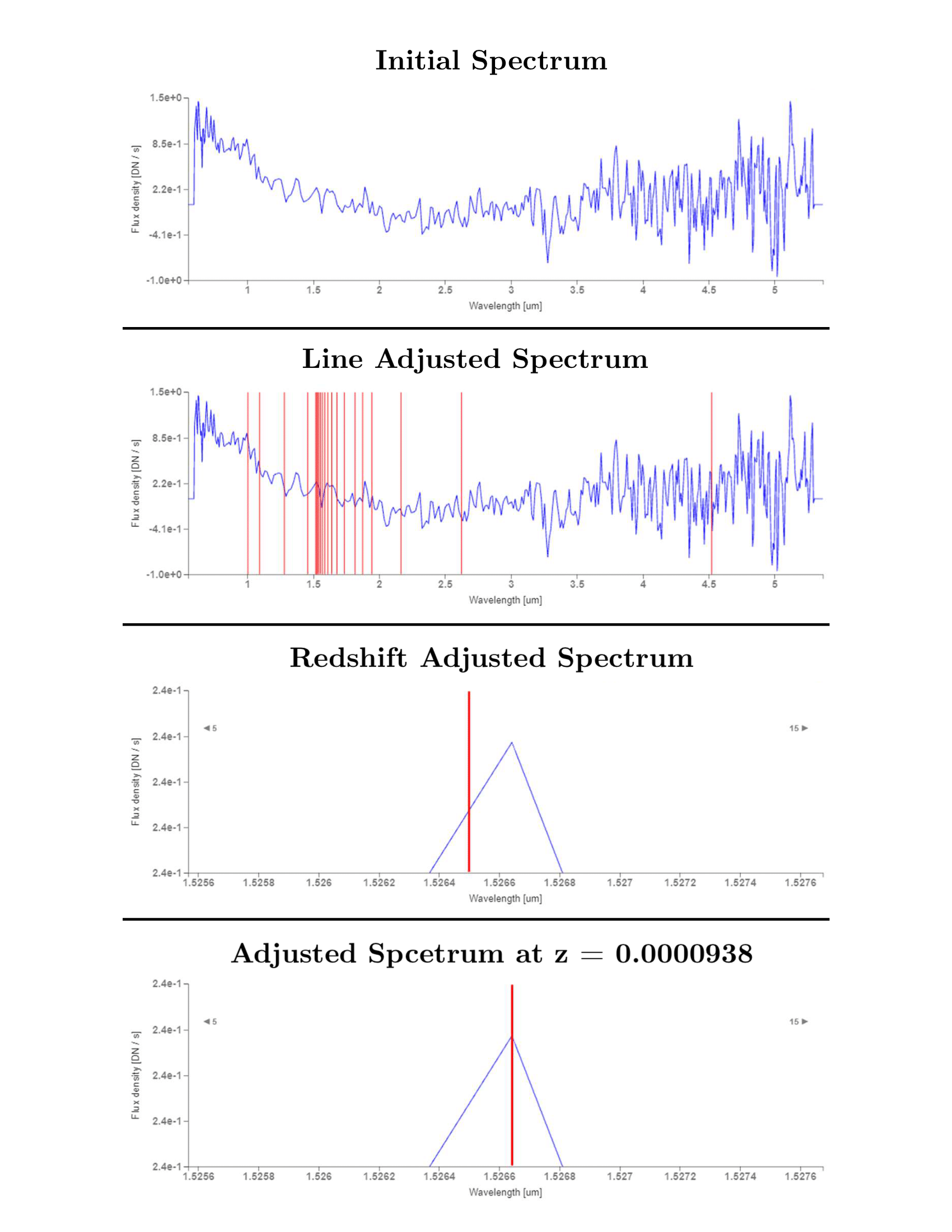}\\
\label{fig:JWST_5}
\end{figure*}

\clearpage


Detailed information for figures \ref{fig:JWST_1}, \ref{fig:JWST_2}, \ref{fig:JWST_3}, \ref{fig:JWST_4} and \ref{fig:JWST_5} can be found in table \ref{tab:JWST_RES1}. The z value was obtained using the equation \ref{eq:z}. The chosen line groups were selected in relation to the wavelength range to be analyzed; the reference line was then determined based on the distinct peaks in the spectrum that were out of phase with respect to the chosen line group.

\begin{deluxetable}{ccccc}[h!]
\tablewidth{0pt}
\tablecaption{Redshift result of the 5 selected regions \label{tab:JWST_RES1}}
\tablehead{
\colhead{ID} & \colhead{Used Lines} & \colhead{Reference Line} & \colhead{$\lambda$ ($\mu$m)} & \colhead{$z$}
}
\startdata
JWST-1 & H2-ISO & H2 10-10 S(6) & 13.1313 & -0.000110 \\
JWST-2 & H2-ISO & H2 11-10 S(7) & 4.5513 & 0.000019 \\
JWST-3 & H-PB & HI 17-4 & 1.5443 & -0.000011 \\
JWST-4 & H-PB & HI 20-4 & 1.5196 & -0.000098 \\
JWST-5 & H-PB & HI 19-4 & 1.5265 & 0.000093 \\
\enddata
\tablecomments{Table \ref{tab:JWST_RES1} shows the redshift results for the 5 selected regions, including the lines used, reference line, wavelength ($\lambda$), and redshift value ($z$). H-PB stands for H-Paschen-Brackett.}
\end{deluxetable}

Using equations \ref{eq:Hubble_1} and \ref{eq:Hubble_2}, we can derive equation \ref{eq:Hubble_3} with respect to distance. By deriving this equation we find the Hubble parameter ($H_{0}$), which, thanks to the research of \citet{12_Riess_2023} will be taken as 67.4 $Km$ $s^{-1}$ $Mpc^{-1}$. Finally, table \ref{tab:JWST_RES2} shows the distances obtained by using the redshift in 5 regions of LMC determined by the analysis of the spectrum provided by the JWST.

\begin{equation}
d_{LMC} = \frac{c \times z}{H_{0}}
\label{eq:Hubble_3}
\end{equation}

\begin{deluxetable}{ccc}

\tablewidth{0pt}
\tablecaption{Distance using the $z$ of the 5 selected regions \label{tab:JWST_RES2}}
\tablehead{
\colhead{ID} & \colhead{$z$} & \colhead{$d_{LMC}$ (Kpc)}
}
\startdata
JWST-1 & -0.00011 & 489.2715 \\
JWST-2 & 0.000019 & 84.51053 \\
JWST-3 & -0.000011 & 48.9271 \\
JWST-4 & -0.000098 & 435.8964 \\
JWST-5 & -0.000093 & 413.6568 \\
\enddata
\tablecomments{Table \ref{tab:JWST_RES2} shows the distance to the Large Magellanic Cloud ($d_{LMC}$) for the 5 selected regions, using the redshift values ($z$).}
\end{deluxetable}

\begin{deluxetable}{ccc}
\tablewidth{0pt}
\tablecaption{Distance to LMC using redshift \label{tab:JWST_RES3}}
\tablehead{
\colhead{$d_{LMC}$ (Kpc)} & \colhead{Std. Error (Kpc)} & \colhead{Sys. Error (Kpc)}
}
\startdata
294.4632 & $\pm$ 93.9525 & $\pm$ 2.1844 \\
\enddata
\tablecomments{Table \ref{tab:JWST_RES3} shows the distance to the Large Magellanic Cloud ($d_{LMC}$) using redshift, along with the associated standard and systematic errors.}
\end{deluxetable}

As shown in table \ref{tab:JWST_RES2}, quite dispersed values are obtained; where $\sigma$ = 210.0766 Kpc is obtained. In table \ref{tab:JWST_RES3} the average result is found together with the standard and systematic error.

If we analyze the z values, we find both positive and negative values; the proper rotation of the galaxy must be taken into account since this influences the recorded z value. In addition, the use of this method is conditioned by the value of $H_{0}$ which, due to its approach, is not a constant and presents uncertainties that are too high.

\subsection{Celestial Mechanics}
Using celestial mechanics there are different methods to determine the distance of a body to the center of mass of the system. One of these methods arises thanks to \citet{Kepler_1964}, where knowing orbital parameters such as its mass and rotation period, the semi-major axis of the orbit can be determined; for this we clear this distance from the equation \ref{eq:Kepler_1} and obtain $a$ using the equation \ref{eq:Kepler_2}.

\begin{equation}
P^{2} = \frac{4\pi^{2}a^{3}}{GM}
\label{eq:Kepler_1}
\end{equation}
\begin{equation}
a = \left ( \frac{T^2GM}{4\pi^2} \right )^\frac{1}{3}
\label{eq:Kepler_2}
\end{equation}

To solve the equation \ref{eq:Kepler_2}, M represents the sum of the masses of the system, which are $M_{MW}$ + $M_{LMC}$, where $M_{LMC}$ = $8.7$ $\pm4.3\times10^{9}$ $M_{\odot}$ and $M_{MW}$ = $4.3\times10^{11}$ $M_{\odot}$; as well as P, which represents the orbital period which is estimated to be around $1.5\times10^9$ years \citep{79_Marel_2002}. Thus, we obtain that $a$ = $48.275$ $\pm0.1575$ Kpc.

\begin{figure*}[h!]
\centering
\caption{Geometric analysis of the LMC's position relative to Earth. Figure \ref{fig:geo} shows the position of the Galactic Center (GC), LMC, and Earth}
\includegraphics[width=6in]{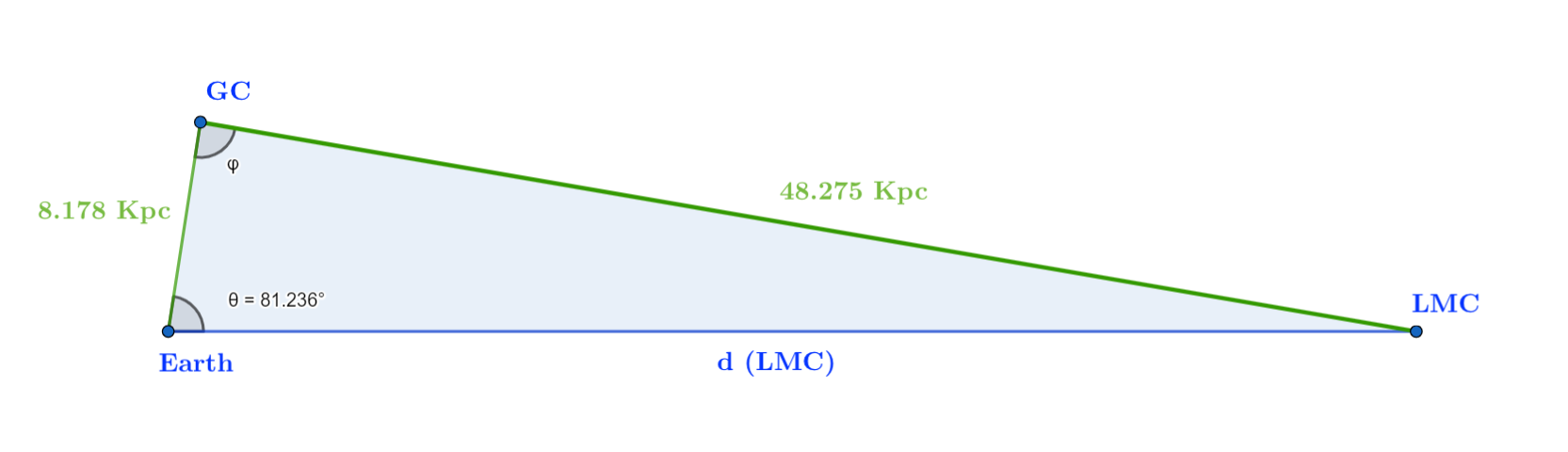}\\
\label{fig:geo}
\end{figure*}

Taking into account that this value of $a$ represents the distance taking into account the center of mass of both bodies, we have to perform a geometric analysis to determine the distance to the Earth, in this investigation the approximation is made that LMC describes a circular orbit or $e$ $\approx 0$. First, the angular separation between the galactic center ($\alpha_{GC}$, $\delta_{GC}$) and LMC ($\alpha_{LMC}, \delta_{LMC}$) must be determined by using the law of cosines, in this case of spherical trigonometry, which is described in the equation \ref{eq:Trig_Esph} with the information from \citet{SIMBAD} in the table \ref{tab:Coordinates}.

\begin{deluxetable*}{ccc}
\tablewidth{0pt}
\tablecaption{CG and LMC coordinates \label{tab:Coordinates}}
\tablehead{
\colhead{Object} & \colhead{$\alpha_{J2000}$ (hh:mm:ss)} & \colhead{$\delta_{J2000}$ (deg:mm:ss)}
}
\startdata
GC & $17^{h}$ $45^{m}$ $39.60^{s}$ & $-29^{\circ}$ $00^{m}$ $22.00^{s}$ \\
LMC & $05^{h}$ $23^{m}$ $34.60^{s}$ & $-69^{\circ}$ $45^{m}$ $22.00^{s}$ \\
\enddata
\tablecomments{Table \ref{tab:Coordinates} presents the coordinates of the Galactic Center (GC) and the Large Magellanic Cloud (LMC) in the J2000 coordinate system.}
\end{deluxetable*}

\begin{equation}
\begin{split}
\theta &= \arccos \left[\sin(\delta _{LMC})\cdot \sin(\delta _{GC}) \right. \\
&\quad + \left. \cos(\delta _{LMC})\cdot \cos(\delta _{GC})\cdot \cos(\alpha _{GC}-\alpha _{LMC})\right]
\end{split}
\label{eq:Trig_Esph}
\end{equation}

This gives an angular separation of $\theta$ = $81.23^{\circ}$. In figure \ref{fig:geo} we observe the angular separations previously obtained together with $d_{LMC}$, the value of $a$ and $d_{\oplus - GC}$ equivalent to $8.178$ Kpc \citep{3_Abuter_2019}.

Finally, by using the laws of sines as described in the equation \ref{eq:geo_eq} indicating the uncertainties of each measurement taken, $d_{LMC}$ is determined obtaining a value of $48.6957$ Kpc, as can be seen in more detail in table \ref{tab:Kepler_RES}.

\begin{equation}
\phi= \arcsin\left( \frac{\sin (82.23^\circ)}{48.275 \pm 0.1575\: \,Kpc} \cdot 8.178 \pm0.175 \: \,Kpc \right)
\end{equation}

\begin{equation}
d_{LMC} = \sin \left( 180^\circ-82.23^\circ-\phi \right) \cdot \: \, \frac {48.275 \pm 0.1575\: \,Kpc}{\sin (82.23^\circ)}
\label{eq:geo_eq}
\end{equation}

\begin{deluxetable}{ccc}
\tablewidth{0pt}
\tablecaption{Distance to LMC using Celestial Mechanics \label{tab:Kepler_RES}}
\tablehead{
\colhead{\textbf{$d_{LMC}$}} & \colhead{Std. Error} & \colhead{Sys. Error} \\
\colhead{Kpc} & \colhead{Kpc} & \colhead{Kpc}
}
\startdata
48.275 & $\pm$ 0.0922 & $\pm$ 0.1598 \\
\enddata
\end{deluxetable}

\section{Results}
By collecting the 6 methods previously described, their results are summarized in table \ref{tab:Results_END}. From this, we can discard the methods that show a considerable deviation. Therefore, we obtain a $d_{LMC}$ = $50.4802$ Kpc recorded in table \ref{tab:END_2}.

\begin{deluxetable}{cccc}
\tablecaption{Methods used to determine the distance to LMC \label{tab:Results_END}}
\tablehead{
\colhead{Method} & \colhead{$d_{LMC}$} & \colhead{Std. Error} & \colhead{Sys. Error} \\
\colhead{} & \colhead{Kpc} & \colhead{Kpc} & \colhead{Kpc}
}
\startdata
Parallax & 41.2459 & $\pm$ 3.5611 & $\pm$ 0.3991 \\
RR Lyrae & 49.9979 & $\pm$ 0.0503 & $\pm$ 0.0481 \\
Classical Cepheids & 50.4930 & $\pm$ 0.1411 & $\pm$ 0.0002 \\
Redshift & 294.4632 & $\pm$ 93.9525 & $\pm$ 2.1844 \\
Celestial Mechanics & 48.2750 & $\pm$ 0.0922 & $\pm$ 0.1598 \\
\enddata
\end{deluxetable}

For the result obtained in table \ref{tab:END_2}, the methods that due to their approach, procedure, or result present a significant deviation with respect to the accepted value of $d_{LMC}$ were discarded. The discarded methods were: Parallax, despite having a relatively close distance to the accepted value, because the standard error presented is considerable and the data range of $\pm0.5\sigma$ that was used must be taken into account, in part, thanks to the large dispersion presented by the collected data. The redshift method was discarded due to the large deviation it presents with respect to the obtained mean; likewise, as previously mentioned, mechanical parameters of LMC such as its rotation speed and inclination must be taken into account to obtain an accurate $z$ value. The celestial mechanics method, despite having a value deviated only by $\approx2$ Kpc, must consider the lack of literature corresponding to the orbital parameters of LMC, in terms of its orbital period or eccentricity.

\begin{deluxetable}{ccc}
\tablecaption{Result Distance to LMC \label{tab:END_2}}
\tablehead{
\colhead{$d_{LMC}$} & \colhead{Std. Error} & \colhead{Sys. Error} \\
\colhead{Kpc} & \colhead{Kpc} & \colhead{Kpc}
}
\startdata
50.4802 & $\pm$ 0.0638 & $\pm$ 0.1377 \\
\enddata
\end{deluxetable}

\section{Discussion}
 Taking into account what is presented in figure \ref{fig:distanceschrono}, we can locate our determined value and it is evident that it follows the trend of a value close to 50 Kpc. Likewise, the parallax method was discarded, this agrees with what was proposed by \citet{47_Reiche_2022} where we see that the parallax method is only effective for nearby stars; added to the uncertainty of $mas$ provided by GAIA according to the magnitude of the star is close to $\approx20 mag$.

The result obtained agrees with the value that is taken as a reference $d_{LMC}$, provided by the extensive history of research and refinement of the \citet{5_Pietrzyński_2013} procedure, through the use of OGLE-III and eclipsing binaries. The use of the JWST and its \textit{NIRSpec} and \textit{MIRI} instruments is highlighted, from which spectra in 5 LMC zones were collected; where finally, the redshift method was discarded for multiple reasons, among them the changing value of $H_{0}$, which, in this research, was used as provided by \citet{12_Riess_2023} who made use of JWST to calibrate this parameter. In the future, new projects such as the ESO ELT, with greater resolving power, are expected to enable distance determinations using geometric methods, which may offer greater precision and extend the scale at which parallax remains effective.

\section{Conclusions}
In this project the distance to the Large Magellanic Cloud was determined using six methods: parallax, RR Lyrae variables, classical Cepheid variables, redshift, celestial mechanics.

For each method used, the standard error and systematic error were calculated and are recorded in table \ref{tab:Results_END}. Of these parameters, the standard error of the parallax method stands out; as well as the standard and systematic error of the redshift method. These present a high value due to their own approach; in the case of parallax, this error is due to the large dispersion of data that was presented, which is why the star record had to be cut to only $\pm 0.5 \sigma$. When talking about the redshift method, it is suggested that the recorded $z$ values are deviated from their current value by not taking into account mechanical parameters of LMC; taking into account these data would present a recurring problem in which the distance is essential to determine the rotation speed in the position of the star that is trying to be recorded.

Likewise, in table \ref{tab:Results_END}, we obtain $d_{LMC}$ using the six methods described. It is observed that the methods used come close by forming an average of approximately 48 Kpc; taking into account only the value of $d_{LMC}$, the distance by the redshift method is discarded, which deviates from this average by more than 240 Kpc.

Finally, when comparing each method used to determine the distance to LMC, it is observed that the parallax and redshift methods present a high value in their standard error; likewise, the values of $d_{LMC}$ obtained, discarding the value using the redshift method, present a standard deviation of $\sigma$ = $3.5893$ Kpc, being relatively close. Therefore, the combined use of the variable star methods is recommended: RR Lyrae or classical Cepheids, or supernovae to determine distances to celestial bodies in the local group; taking into account that they have a great precision and in the case of LMC and SMC there is a large record of the region, which translates into a large availability of data. In this way, each method can be complemented, reducing their uncertainties and obtaining a value with greater precision; either due to the morphology of the galaxy itself and the distribution of variable stars, or due to the record of different supernovae in it.

The distance to the LMC, $d_{LMC}$ = $50.4802$, is therefore determined using RR Lyrae variables, classical Cepheids, and the supernova SN1987a. This value is $+0.5102$ Kpc away from the accepted value of this distance, obtained by \citet{26_Pietrzyński_2014} using EBs with a value of $d_{LMC}$ = $49.97$ Kpc and an uncertainty of 2$\%$.

Considering that OGLE-IV and JWST will be conducting data capture and collection in this region of interest in the near future, this research project will be able to continue, allowing for an expansion in data collection for greater precision in the different variables that are analyzed in this region.

\section{Data Availability}
Distance data and observational data were cited above. Also, according to the suggestions made by \citet{9_Chen_2022}, the necessary data are available to recreate the tables and figures available in the repository \url{https://github.com/jj-sm/DistanceToLMC}.

\section{Code Availability}
The analysis of this research was performed in Python 3.9.13 and makes use of open source libraries, including Numpy 1.26.1, Astropy 5.3.4; Astroquery 0.4.6; Matplotlib 3.8.0. Likewise, the software JS9 \citep{JS9}., Stellarium \citep{Stellarium_2021}. and Jdaviz \citep{MAST} were used. Full access to the code and scripts used are available in the repository \url{https://github.com/jj-sm/DistanceToLMC}.

\newpage

\bibliography{ref}{}
\bibliographystyle{aasjournalv7}



\end{document}